\documentclass[fleqn,10pt]{style/wlscirep}
\usepackage[utf8]{inputenc}
\usepackage[T1]{fontenc}
\usepackage{amsmath}
\usepackage{multicol}
\usepackage{multirow}
\usepackage[none]{hyphenat}
\usepackage{longtable}
\usepackage{makecell}
\usepackage{hyperref}
\hypersetup{colorlinks=true, citecolor=blue}
\usepackage{graphicx}
\usepackage[font={small,sf},labelfont=bf]{caption}
\usepackage{subcaption}
\graphicspath{{./figures/}}

\title{Mobility-based contact exposure explains the disparity of spread of COVID-19 in urban neighborhoods}

\author[1]{Rajat Verma}
\author[1]{Takahiro Yabe}
\author[1,*]{Satish V. Ukkusuri}
\affil[1]{Lyles School of Civil Engineering, Purdue University, West Lafayette, 47906, U.S.A.}
\affil[*]{Corresponding author: sukkusur@purdue.edu}

\begin{abstract}
The rapid early spread of COVID-19 in the U.S. was experienced very differently by different socioeconomic groups and business industries.
In this study, we study aggregate mobility patterns of New York City and Chicago to identify the relationship between the amount of interpersonal contact between people in urban neighborhoods and the disparity in the growth of positive cases among these groups.
We introduce an aggregate \textit{Contact Exposure Index} (CEI) to measure exposure due to this interpersonal contact and combine it with social distancing metrics to show its effect on positive case growth.
With the help of structural equations modeling, we find that the effect of exposure on case growth was consistently positive and that it remained consistently higher in lower-income neighborhoods, suggesting a causal path of income on case growth via contact exposure.
Using the CEI, schools and restaurants are identified as high-exposure industries, and the estimation suggests that implementing specific mobility restrictions on these point-of-interest categories are most effective.
This analysis can be useful in providing insights for government officials targeting specific population groups and businesses to reduce infection spread as reopening efforts continue to expand across the nation.
\end{abstract}

\begin{document}

\flushbottom
\maketitle
\thispagestyle{empty}

\section*{Introduction}

The rapid worldwide spread of the novel coronavirus disease (COVID-19) has caused significant distress to citizens and governments worldwide and has warranted unprecedented control measures on human mobility such as travel bans, mandatory quarantines, and stay-at-home/shelter-in-place orders.
In the United States, like many other countries, these restrictive policies have incurred huge economic losses \cite{Thunstrom2020}, leading to experts showing interest in targeted policy-making based on differential effects of the virus on different subject and activity types.
These policies may be either supportive, such as the CARES Act relief based on household income \cite{USCongress2020}, or restrictive, such as heightened restrictions on some business activities more than others, such as schools, bars, concerts, sports events, and indoor dining.
Similarly, in some cases, specific neighborhoods within cities may be classified as ``coronavirus hotspots'' that are subjected to additional precautions and heightened restrictions, such as the New Rochelle community in the NYC metropolitan region in March 2020 \cite{Nir2020}.

It has been shown that these policies have had differential impacts on the livelihood and health of different socioeconomic groups within cities and states, with certain factors such as income \cite{Weill2020}, education level \cite{Brough2020}, and race \cite{Jia2020, Hooper2020} standing out as the principal discriminants, in addition to age which is inherently strongly correlated with higher susceptibility to infection by and morbidity due to the coronavirus.
In considering targeted policies, such as targeting urban neighborhoods, it then becomes crucial to understand how these different social groups may respond to such policies in terms of mobility and the growth of the disease.

A crucial hindrance to this approach is a lack of high-fidelity publicly available epidemiological data related to the disease.
State governments and popular COVID-19 trackers such as the Covid Tracking Project \cite{CovidTrackingProject} and the JHU Coronavirus Research Center \cite{Dong2020} generally provide the data at the county level.
Epidemiological data are scarce for smaller spatial units such as zip code tabulation areas (ZCTAs) and have been only recently made publicly accessible and by only a few regional governments such as those of NYC \cite{nychealth}, Illinois \cite{idphhealth}, and Ohio \cite{ohiohealth}.

Many studies that have exploited county-level data have shown evidence of the effectiveness of these social distancing measures and related mobility reduction in reducing the rate of daily new COVID-19 positive cases, both in the U.S. \cite{Courtemanche2020, Abouk2020, Andersen2020} as well as in other countries \cite{Gatto2020, Lai2020}.
However, there is one major concern associated with the current research focusing on the relationship between aggregate mobility and the spread of COVID-19 that we address in this study.
In the absence of human movement data at the individual scale, most studies consider mobility as an aggregate entity measured by the total number of trips between counties or other large regions \cite{Badr2020, Kraemer2020, Linka2020} or distance-based measures \cite{Yabe2020}.
Although overall movement traffic flow is a reasonable proxy of social distancing among travelers as a whole between regions, it suffers from two crucial limitations.
First, social distancing inherently involves physical interaction between individuals which is not directly captured by traffic flow and distance traveled.
Travelers who travel long distances alone or in small groups with little exposure to contact with others, such as in private vehicles, are misclassified as equally potential vectors of the virus as those who travel in close proximity with large groups, such as public transit during the peak hours \cite{Labonte-Lemoyne2020}.
Second, such measures may not be a good indicator of the contact exposure travelers are subjected to at their destinations, where they are likely to spend more time in proximity of other visitors than the means of travel.
This includes the number and physical spacing of other visitors at the destination as well as their common dwell time in contact with each other.
These are important factors to be taken into account since the Centers for Disease Control and Prevention specifically highlights via its famous 6-feet-15-minutes rule for a close contact \cite{CDC2020}.
While recent studies have considered the distribution of trips by dwell time \cite{Huang2020} and crowd density based on non-residential square footage \cite{Wang2020PPR:PPR242642} at certain trip destinations, these have been studied in isolation as time series trends but not as a comprehensive measure of exposure.

In this work, we address these two limitations - lack of a contact exposure-based mobility measure, and mobility analysis at the ZCTA level - in understanding the differential impact of mobility restrictions on socioeconomic groups within cities and states.
We leverage anonymized origin-destination foot traffic movement patterns based on mobile phone GPS records and aggregated by SafeGraph Inc. at the level of urban neighborhoods (as ZCTAs) and the business category/industry of the trip destinations, hereby referred to as places of interest (POIs).
Although there are significant concerns with using mobile phone GPS data, such as low penetration rate and privacy and data protection issues \cite{Oliver2020, DeMontjoye2018}, they have been nonetheless used both in general mobility analysis \cite{Sia-Nowicka2016, Vazquez-Prokopec2013} as well as specifically for COVID-19 \cite{Abouk2020, Andersen2020}.

Based on the insights obtained from the studies exploiting county-level trip/distance measures, we hypothesize that the number of positive coronavirus cases in residential neighborhoods is positively correlated with the amount of contact exposure their residents are subjected to when they travel outside their neighborhoods.
Furthermore, we hypothesize that this difference in aggregate contact exposure emanates from, among other things, the difference in several socioeconomic factors of the residents.
We test this causal pathway using structured equations models based on a panel of daily mobility and case growth data of ZCTAs in New York City (NYC) and the Chicago metropolitan area up to June 2020.
Finally, we analyze the contribution of exposure to 12 major trip destination types (e.g., schools, hospitals, and restaurants) to the overall exposure to understand which industries could be targeted for tighter mobility restrictions in the event of a growth spurt in COVID-19 cases.
In doing so, we also study the variation in the negative effect of income on exposure felt in these industries.

\section*{Results}

\subsection*{Lower income groups had higher exposure and more cases}

\begin{figure}[h]
    \centering
    \includegraphics[width=\textwidth]{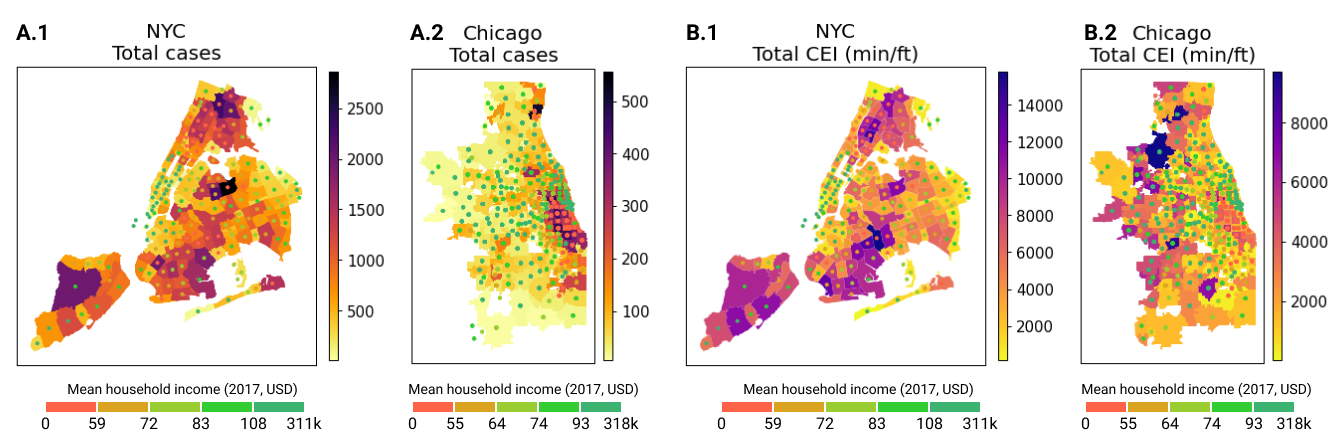}
    \caption{Map of study cities showing total cases and CEI.
    ZCTAs of the two study cities showing (\textbf{A}) total cases and (\textbf{B}) total contact exposure index (min/ft) during 20-26 April. For reference, the quintile class of the mean household income of the ZCTAs are also shown as colored centroid dots.}
    \label{fig: city map}
\end{figure}

In this section, we analyze the heterogeneity of infection spread by socioeconomic status in NYC and Chicago which had been some of the worst-hit cities by COVID-19 during the first wave of the pandemic \cite{Glaeser2020much}.
We quantify this heterogeneity along the dimension of mean household income of the ZCTAs.
We also use an unweighted gravity model-like approach to distribute the exposure (CEI) from each POI to the ZCTA of its visitors on each day and then aggregate over all POIs to get the exposure level of each ZCTA.
Fig. \ref{fig: city map} shows the cumulative number of positive cases and the cumulative CEI of the two study cities in the week of 20-26 April, since the earliest date for which reliable ZCTA-level epidemiological data are available for Chicago is April 18.
The dots at the centroids of the ZCTAs depict their income class, given by the quintile of the average household income distribution of the ZCTAs as of 2017.

\begin{figure}[ht]
    \centering
    \includegraphics[width=\textwidth]{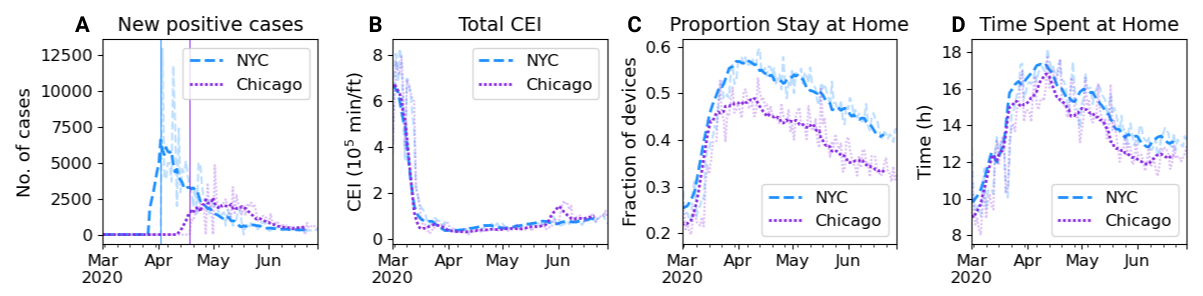}
    \caption{Trends of cases and mobility measures.
    Daily variation of (\textbf{A}) new positive cases, (\textbf{B}) contact exposure index, (\textbf{C}) fraction of devices registered as staying at home all day, (\textbf{D}) median time spent at home.
    The light shaded curves denote the daily trends while the dark ones depict their 7-day forward-shifted moving average.
    The vertical lines in panel \textbf{A} represent the first dates of available data of the number of cases.}
    \label{fig: city-level trends}
\end{figure}

At first glance, it can be observed that lower income regions (such as The Bronx in NYC and southern Chicago) had disproportionately more positive cases than their higher income counterparts in this week.
They also experienced higher exposure to contact as measured by total CEI during that period.
While cases were relatively more evenly distributed in NYC, with some peaks in neighborhoods in Queens and King Counties, they were much more concentrated in South and downtown Chicago, albeit much fewer than NYC in general.
Moreover, the virus transmission started to decline in Chicago about two weeks after this week, while it had already started declining since at least early April (see Fig. \ref{fig: city-level trends}A).
On the contrary, though, mobility began increasing in April in both the cities after the initial phase of mobility plummet following the issuance of stay-at-home orders on March 20 in NYC \cite{sahnyc} and Chicago \cite{sahchicago}, with total exposure to contact initially decreasing rapidly (Fig. \ref{fig: city-level trends}B) and more people spending more time at home (Fig. \ref{fig: city-level trends}C and D).

\begin{figure}[h]
    \centering
    \includegraphics[width=0.7\textwidth]{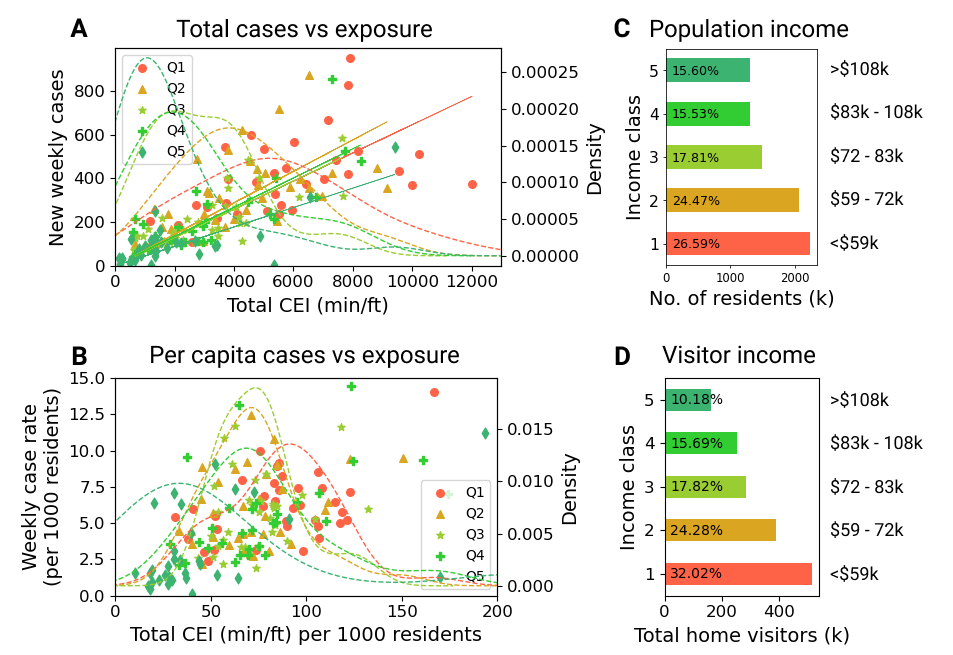}
    \caption{Relationship between cases, exposure, and income.
    (\textbf{A}) New positive cases versus total exposure (CEI) in the ZCTAs of NYC in the week of 6-12 April, differentiated by the ZTCA income quintiles, along with the probability density distribution of ZCTAs by income class; (\textbf{B}) same as \textbf{A} but with values divided by ZCTA population. For reference, distributions of (\textbf{C}) population of ZCTAs and (\textbf{D}) the number of visitors in this week across these income classes are also shown, along with the intervals covered by these classes.}
    \label{fig: cei vs cases snapshot}
\end{figure}

On closer inspection, the relationship between exposure-based mobility (measured by CEI), the spread of the virus, and income becomes more evident when looking at the total exposure subjected to the population of neighborhoods (see Fig. \ref{fig: cei vs cases snapshot}A).
However, even after controlling for the population of the neighborhoods, we find that exposure per capita is strongly associated with cases per capita (Fig. \ref{fig: cei vs cases snapshot}B), although the strength of this correlation reduces after controlling for population.
For reference, the distribution of population of the ZCTAs across these income groups is shown in Fig. \ref{fig: cei vs cases snapshot}C.
This effect likely occurs due to a higher proportion of visitors belonging to lower-income neighborhoods after the imposition of the stay-at-home order in NYC (Fig. \ref{fig: cei vs cases snapshot}D).
This observation supports the idea that lower income people were more susceptible to infection after the stay-at-home orders came into effect in these cities primarily because of the nature of their professions, mainly a higher representation of jobs requiring on-site work and/or belonging to essential services such as nursing, grocery store operations, etc. \cite{Weill2020}

\subsubsection*{Econometric Modeling}

We hypothesize that the difference in the caseload of lower-income neighborhoods can be explained by the difference in the amount of exposure to contact they were subjected to during the early phase of the lockdown.
We highlight the importance of measuring this exposure with CEI and other social distancing metrics instead of relying on the variation of number of trips since the latter may misrepresent the exposure by unnecessarily counting solo trips and discounting the interaction with others at the destination.

Here, we test this hypothesis by testing the strength of the causal path of household income to virus transmission through a latent measure of exposure to social contact using a structured equations model (SEM).
We specify this exposure measure to be latent so that it can take into account the effects of social interaction at places of commercial as well as non-commercial activity and compensate for the shortcomings of the used mobility variables.
While exposure at commercial places is captured reasonably well with CEI at POIs (which include most major places of commercial activity except offices of private firms), the two social distancing measures - \textit{Prop Home} and \textit{Time Home}, estimate the exposure due to travel outside home without considering social interaction.

We develop daily SEMs where for each day $t$, a causal pathway is assumed from 6 static socioeconomic variables, particularly mean household income, to the daily number of new cases via latent exposure measured by the daily mobility variables (mutually correlated).
Models of similar design have been used to show the impact of inter-county travel flow on case growth rate \cite{Xiong2020}, though without considering contact exposure as a facilitator.
For more details of the model structure used, see the Materials and Methods section.
The variables are described in Table \ref{table: variables used in SEM}.

\begin{table}[h]
\centering
\fontsize{9}{11}\selectfont
\sffamily
\caption{\textbf{Description of SEM variables.} Description of the independent, target, and latent variables used in the daily structured equations models, along with their symbols and ranges.}
\begin{tabular}{l l p{0.5\textwidth} c }
    \hline
    \textbf{Category} &
    \textbf{Name} &
    \textbf{Description} &
    \textbf{Range} \\
    \hline
    \multirow[t]{6}{6em}{Socioeconomic, $\mathbf{S}_i$} & 
    \textit{Income} & Natural logarithm of mean household income of ZCTA $i$ & $[0, \infty)$ \\
    & \textit{Low Edu} & Fraction of population having a high school diploma or less & $[0,1]$ \\
    & \textit{Poor} & Fraction of population classified as living below the poverty line & $[0,1]$ \\
    & Age & Fraction of population aged 65 years and above & $[0,1]$ \\
    & \textit{Black} & Fraction of population which identifies as African American (monoracial) & $[0,1]$ \\
    & \textit{Transit} & Fraction of population whose primary mode of commute to work is public transit (buses, monorail, or subway) & $[0,1]$ \\
    \hline
    \multirow[t]{3}{6em}{Mobility, $\mathbf{M}_{i,w}$}
    & \makecell[tl]{\textit{CEI} \\ $E_{i,w}$} & Weighted total contact exposure index (min/ft) of POI visitors living in ZCTA $i$ in the time window $w=[t-7,t)$, transformed with the mapper $f(x)=\ln(1+x)$ & $[0,\infty)$ \\
    & \makecell[tl]{\textit{Time Home} \\ $T_{i,w}$} & Fraction of day spent by the mobile devices within the Geohash7 of their owners's home in ZCTA $i$ on day $t$, averaged over the window $w$ & [0,1] \\
    & \makecell[tl]{\textit{Prop Home} \\ $P_{i,w}$} & Proportion of mobile devices registered as staying within the home Geohash7 of residents of ZCTA $i$ all day on at least one day in the window $w$ & [0,1] \\
    & \makecell[tl]{\textit{Devices} \\ $N_{i,w}$ \\ \small{(not used)}} & Number of mobile devices belonging to residents of ZCTA $i$ on day $t$; not used directly in the model but only to aggregate \textit{Prop Home} and \textit{Time Home} over the window $w$ & $[0, \infty)$ \\
    \hline
    \multirow[t]{2}{6em}{Health}
    & \makecell[tl]{\textit{Cases} \\ $y_{i,t}$} & Number of new positive cases in ZCTA $i$ on day $t$, also transformed with $f(x)$ & $[0,\infty)$ \\
    \hline
\end{tabular}
\label{table: variables used in SEM}
\end{table}

\begin{figure}
    \centering
    \includegraphics[width=0.8\textwidth]{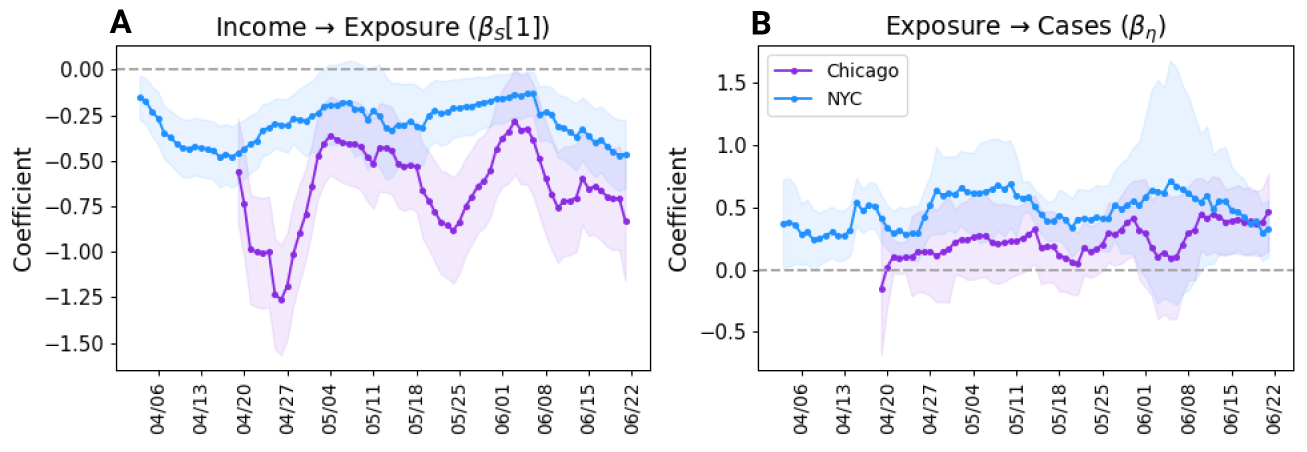}
    \caption{Estimates of important parameters of daily SEMs.
    Daily series of coefficient estimates of the two SEM relationships of interest for Chicago and NYC shown as 7-day moving averages: (\textbf{A}) effect of income on exposure, and (\textbf{B}) effect of exposure on new cases.
    The shaded region indicates the region spanned by estimate $\pm$ standard error.}
    \label{fig: sem params}
\end{figure}

The parameter estimates of the two main relationships of interest in the daily SEMs of the two cities are shown in Fig. \ref{fig: sem params}, along with the standard error of these estimates.
The coefficients $\beta_S[1]$ measuring the effect of income on exposure (panel A) on exposure are consistently negative for both the cities.
This implies that residents of lower income neighborhoods in these cities remained more likely to coming in close contact with other individuals throughout the study period.
This effect is higher in the case of Chicago, although the large fluctuations in this effect are not correlated with any remarkable shift in other variables, such as lifting of the lockdown or a sudden and brief change in public response to COVID-19 that could have triggered this change.
This difference in effects between NYC and Chicago could also emanate from the already large pool of infected people and higher mobilization of resources in NYC due to its uniquely intense peak of cases in mid-March \cite{goyal2020clinical}.

A similar consistency is also observed in the effect of exposure on daily new cases where an increase in exposure is linked with a corresponding increase in the number of cases (Fig. \ref{fig: sem params}B).
In this case, however, the effect is higher in the case of NYC compared to Chicago, implying that NYC was more sensitive to mobility changes in terms of virus transmission than Chicago during the initial months after lockdown.
After controlling for unobserved variables in these models, one could interpret this in this way - even if household income in NYC is not as considerable an indicator of exposure to contact as in Chicago, contact exposure contributed more substantially to the growth of cases in NYC than in Chicago.

These observations provide the core insight for understanding the causal mechanism of trip destination-based contact exposure on the course of COVID-19 in the early phase of lock-downs in these cities.
In the next section, we discuss how this exposure varied across different destination types, which could be used to identify the industries active in helping spread COVID-19 faster.

\subsection*{Exposure by Destination Types}

To better understand the characteristics of trips that contribute more to exposure to contact, we next discuss the travel trends in NYC and Chicago to POIs of different industries.
The publicly available Google Community Mobility Report\cite{googlemobility} has been commonly used to study the differences in travel behavior across trip categories \cite{Abouk2020, Wang2020}. However, it only provides data at the state or county level and for a select travel categories, such as home, work, groceries, etc., meaning there is limited opportunity to explore specific industries of interest, such as bars and hospitals.
The SafeGraph mobility patterns are provided at the POI level, so they can be used to study these categories in detail, such as in \cite{Jay2020}.

\begin{figure}[h]
    \centering
    \includegraphics[width=\textwidth]{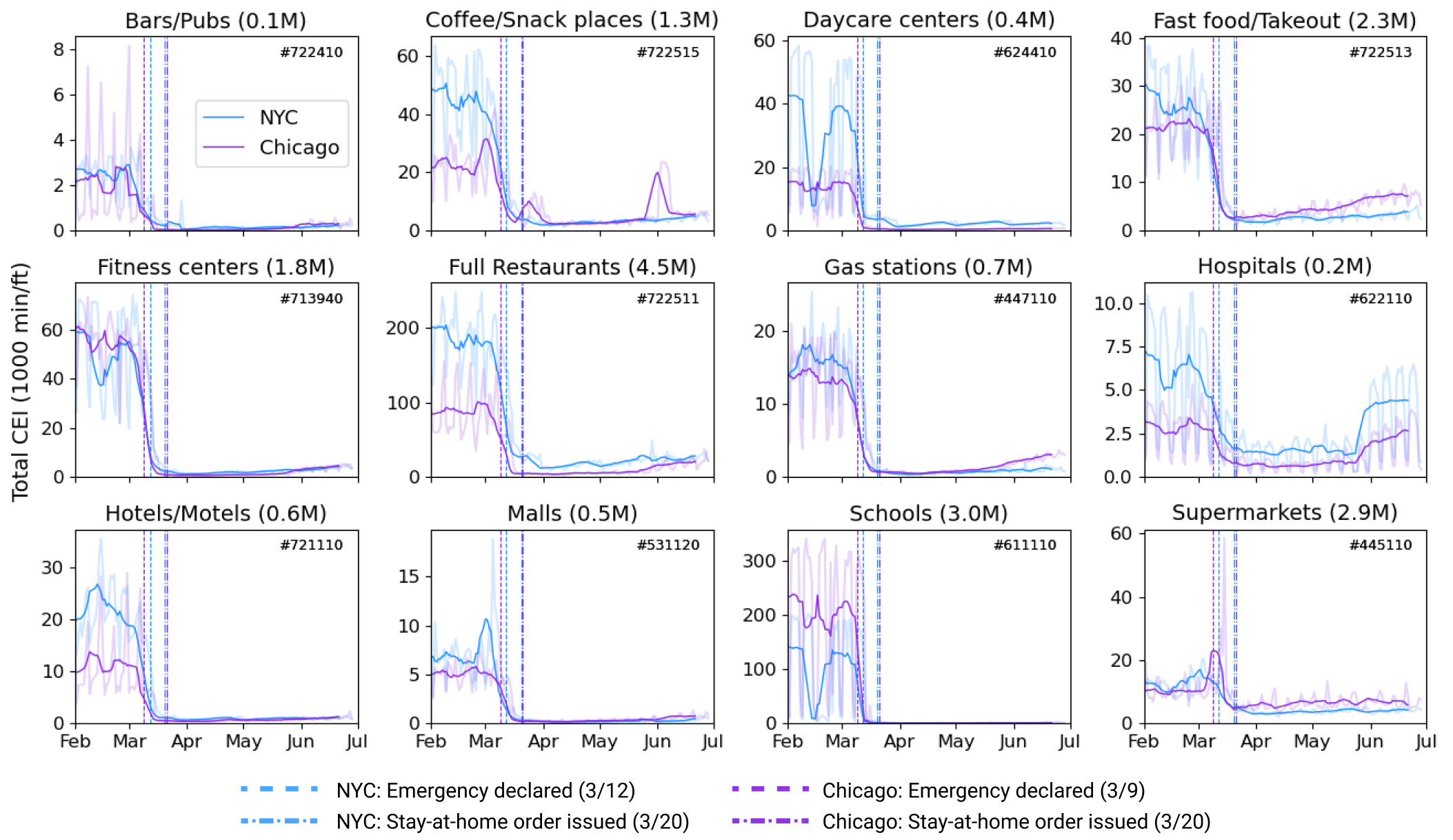}
    \caption{Trends of CEI by industry.
    Daily trends of CEI to POIs of 12 popular industries in NYC and Chicago, shown as 7-day moving averages.
    The lighter shaded curves are the daily trends.
    For reference, the 6-digit NAICS code and the total number of multi-person visits to POIs between February 1 and June 28.
    The dates of two main mobility restrictive policies in these cities are also shown.}
    \label{fig: cei by industry}
\end{figure}

We focus on 12 popular industries (by daily visits) in the study cities and label them according to their industry codes as per the North American Industry Classification System (NAICS).
The trends of the total daily exposure (CEI) across these industries are shown in Fig. \ref{fig: cei by industry}, along with their NAICS codes and total number of visits to their POIs in the entire city (excluding lone hourly visits).
In this figure, it can be seen that all of the industries experienced a drastic decline after the declaration of emergency in the two cities, with many categories falling close to zero exposure, such as schools and malls immediately after the lockdown (stay-at-home rule).
Since then, exposure has increased in hospitals and at fast food places but has largely remained negligible compared to before emergency.
While schools and fitness centers have seen the biggest plummet in exposure, supermarkets have seen the lowest decline after a brief weekend surge in Chicago, likely due to panic buying.

There are a few interesting shifts in travel behavior across the two cities.
CEI in the Chicago metropolitan area had been lower on average than NYC at supermarkets, fast food places, and gas stations prior to the shutdown, but this pattern started reversing afterwards.
This could simply be a consequence of the severity of enforcement of lockdown practices in NYC, with several reports discussing the severe punitive actions being taken against social distancing violators during this period.

\subsubsection*{Contribution of Industries on Exposure}
\begin{figure}[h]
    \centering
    \includegraphics[width=\textwidth]{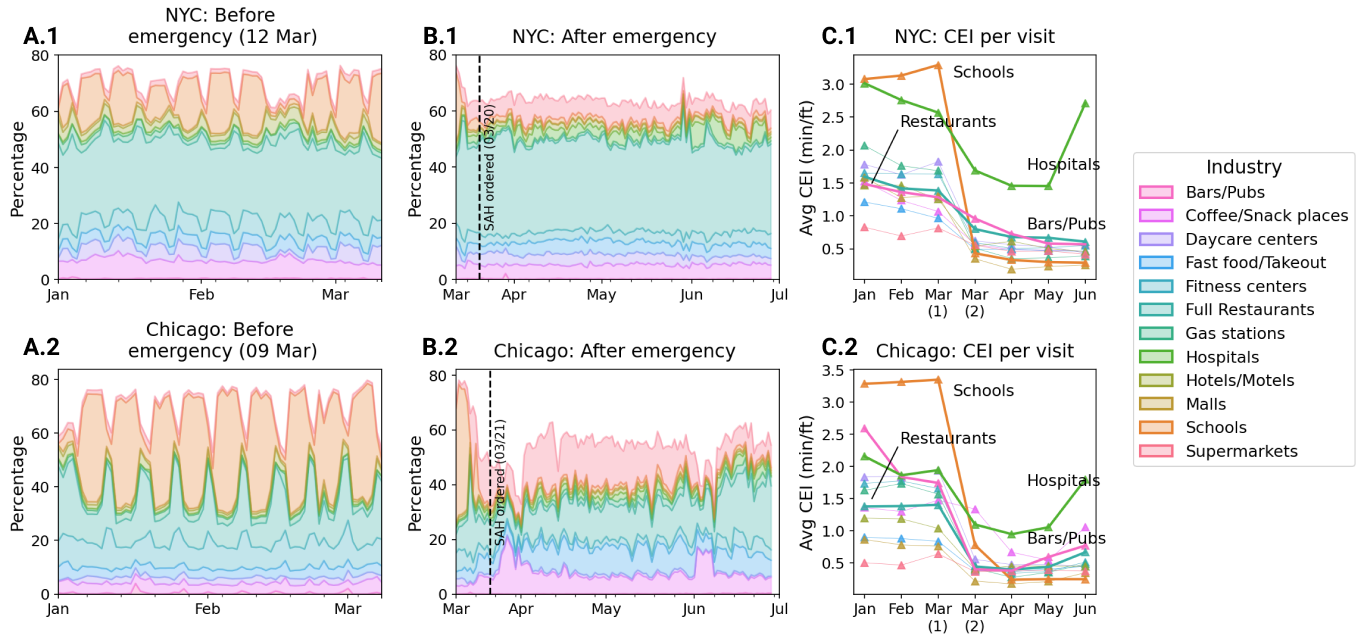}
    \caption{Contribution of CEI by industry. Proportion of total daily CEI attributable to the 12 industries of interest in the study cities (\textbf{A}) before and (\textbf{B}) after the declaration of emergency.
    The other industries contribute the remainder of the CEI, denoted by the empty region in the chart areas.
    (\textbf{C}) Monthly change in average CEI per visit in the two cities, with March divided into two parts (before and after March 15).}
    \label{fig: cei composition by poi}
\end{figure}

Though Fig. \ref{fig: cei by industry} provides an overview of mobility patterns across the major industries considered here, an important factor in the consideration of categorical restrictions is the contribution of these industries to the overall exposure to contact at a macroscopic level.
This is clarified in Fig. \ref{fig: cei composition by poi} where panels A and B show the proportion of CEI coming from POIs in the 12 important industries.
The differences before and after the emergency declaration are evident in many categories.
An interesting shift in the pattern of contact exposure is the near eradication of weekly recurring patterns in both the cities, primarily achieved through the closure of services that typically show strong traffic variation between weekdays and weekends, such as schools, daycare centers, fitness centers, and eating places.

Schools offer a particularly interesting case in point.
Public schools in NYC were closed on March 16, but many private schools and school districts had already started closing a week ago.
Prior to closure, schools exhibited some of the highest exposure, both in terms of total CEI (Fig. \ref{fig: cei composition by poi}A) and average CEI per visit (Fig. \ref{fig: cei composition by poi}C).
This makes sense given that visits to schools typically have much higher dwell time (typically 4-5 hours) than other POIs and also have more visitors in general (see Fig. \ref{fig: cei by industry}).
The drop in exposure following school closure is even starker in Chicago where recurring periods of high CEI vanished almost overnight close to its date of issuance of the stay-at-home order.
These observations support the decision of the public authorities of closing schools on the grounds of exposure to contact.

The full-service restaurant industry also stands out as being the dominant destination type for visitors over the entire study period for both the cities.
Although this industry has not seen a decline in exposure as sharp as schools and daycare centers, it has had the most impact in the total reduction of exposure.
This was likely facilitated by the differing degrees of closure following lockdown, with most restaurants remaining fully shut while a few others provided outdoor dining services \cite{Imbruce2020}.

\begin{figure}[ht]
    \centering
    \includegraphics[width=\textwidth]{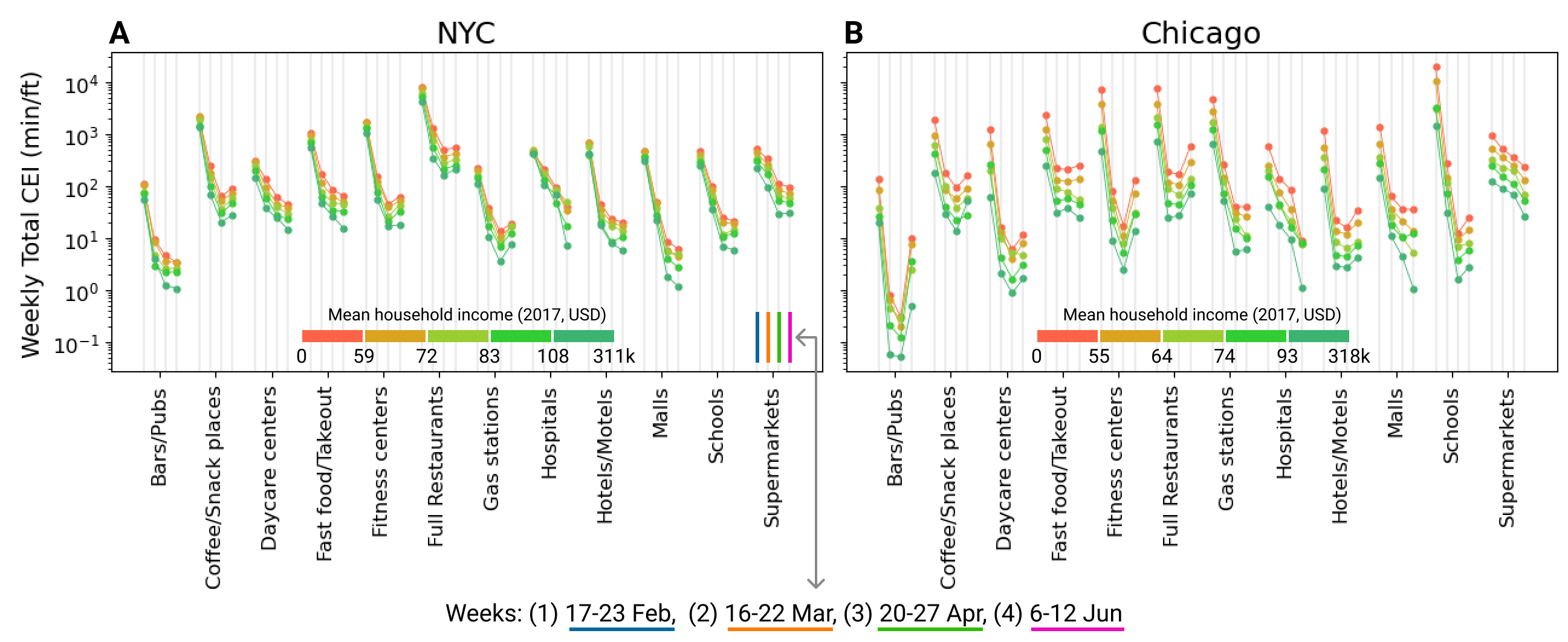}
    \caption{Phase comparison of CEI by industry and income.
    Variation of CEI generated by residents of neighborhoods of 5 income classes by visiting POIs of the industries of interest in 4 specific weeks representing different phases of the study period in (\textbf{A}) NYC and (\textbf{B}) Chicago.}
    \label{fig: cei by income and poi}
\end{figure}

The results for these two industries provide affirmation for the actions taken by public authorities in exercising special restrictions on them.
However, these decisions have had implications on the disparity of contact exposure across neighborhoods of different income levels.
This difference by industry is highlighted in Fig. \ref{fig: cei by income and poi}.
This figure shows the relationship between POI industry and contact exposure considering the income level of the people who visit these POIs and reaffirms the sharp decline in the exposure (CEI) in  schools, fitness centers, and bars, especially in Chicago.
We chose 4 weeks representative of the different phases of mobility restrictions in the study period to summarize the evolution of this disparity - mid-February (pre-lockdown), mid-March (just after lockdown), late April (a month afterwards), and early June (beginning of reopening).

A clear pattern in this figure is a consistent ordering of CEI with respect to income classes across all the categories and both the cities.
Even though CEI declined very sharply after the implementation of stay-at-home, this order did not change much.
Interestingly, the ratio of the exposure generated by the lower-income neighborhoods to that generated by the higher-income ones did not change substantially in Chicago but increased substantially in NYC.
This can be inferred by noting the difference in the width of the bands across the industries at the left end (pre-lockdown) with the right end (lockdown and reopening), which on a linear scale represents the ratio of CEI.
It could be argued that a stricter lockdown in NYC could have triggered a more polarized response from the public partly attributable to the inability of the lower-income people to stay at and work from home.

These observations provide new and confirm already accepted insights pertaining to mobility and spread of COVID-19, such as the rise of socioeconomic disparity in cities during at least the early period of the pandemic and the role of special restrictions on certain types of places.

\section*{Discussion}

Social distancing policies like stay-at-home orders and closure of many services have caused widespread decline in overall mobility since mid-March due to the spread of COVID-19 in the U.S.
While research has shown that these restrictions have been associated with an increase in socioeconomic disparity among urban neighborhoods, little work has been done on understanding the mechanism of this change.
Also, while current research in this respect has often relied on macroscopic mobility measures like population flow and distance travelled, there has been limited work which seeks to understand the effect of mobility by exploiting the knowledge that one of the most important causes of the spread of this disease is coming in close contacts with an infected person.

In this study, we attempt to create the relationship bridge between human mobility involving high exposure to COVID-19 and the effect of the rapid change in this mobility on the rise of socioeconomic disparity in U.S. cities by analyzing the contact exposure-based mobility patterns of Chicago and New York City in the first four months of the widespread outbreak of the pandemic.
Based on aggregate mobile phone-based mobility data provided by SafeGraph Inc., we develop a Contact Exposure Index (CEI).
This is an aggregate mobility metric based on three important factors associated with the idea of socio-physical contact - the total number of people who visit a place within the city, the area over which the visitors are spread over, and the duration of their stay there, with a special consideration of the schedule of their visits.

We observe that income is a consistently strong indicator of contact exposure measured by CEI in both the cities.
Recognizing that CEI does not capture socio-physical interaction at places not classified as places of interest (POIs), we conceptualize an abstract notion of contact exposure measured by CEI in combination with two social distancing metrics that we believe affect contact exposure - proportion of the mobile phone-tracked population staying at home all day, and the amount of time they spend at home.
We establish the negative effect of mean household income of zip code areas on this latent contact exposure and in turn the positive effect of exposure on the number of COVID-19 cases using a time series of structured equations models.

We then attempt to explain the composition of contact exposure by the destination categories (industries) of the trips generating that exposure, measured by CEI.
We observe that heightened restrictions on mobility to POIs of certain categories, such as schools and restaurants, have contributed substantially to the decline in the overall exposure to socio-physical contact in these cities, while the effect of closure of bars has limited contribution to this decline.
This lends support to the idea of industry-specific targeted lockdown policies that the government officials have been implementing throughout the pandemic.
Finally, we also observe that the disparity in contact exposure by income class considerably increased over all of the important industries after lockdown in Chicago but not much in New York.

Given the practical importance of these insights, we also recognize the numerous limitations with this macroscopic approach of quantifying exposure to COVID-19 manifested from close interpersonal contact.
First, the contact exposure index we propose is based on assumptions about the spatiotemporal positioning of visitors within POIs that are highly ideal and likely uncommon.
Second, the scale of this measure is highly dependent on the true number of visitors at POIs which under the currently available data is a valid concern due to issues related to low representative coverage of the mobile devices tracked by SafeGraph.

Having said that, we assert that this measure nevertheless provides more pertinent information about COVID-19 transmission and is more comprehensive than flow and distance-based measures and should be pursued as a tool of monitoring the progress of policies pertaining to mobility restriction, especially now that cities have begun reopening despite the pandemic soaring in the U.S.
We hope to extend this study to provide a sound basis to the validity and practical applications of this measure and the insights in this study in the future.

\section*{Materials and Methods}

\subsection*{Data Description}
\subsubsection*{Mobility}
Two mobility datasets were obtained from SafeGraph Inc. whose foot traffic records have been used in many studies related to mobility during the COVID-19 pandemic \cite{Brough2020, Andersen2020}.
The first dataset provides information about trips originating from different CBGs as defined in the American Community Survey (ACS) of 2013-2017.
It includes relevant information such as the number and duration of devices staying in their home CBGs and their destination CBGs.

The second dataset contains information of about 4 million POIs across the U.S., including their unique 6-digit NAICS code (defining the industry of the POIs), floor area, enclosing CBG, hourly count of trips to these POIs, weekly distributions of trip distance, weekly trip dwell time, and origin CBGs of the visitors.
We excluded the POIs located inside hospitals (mostly fast food restaurants) because of classification error due to the surge in visits to hospitals in the study cities during the peak of the pandemic.

\subsubsection*{Epidemiological Data}
Information about daily new tests, positive cases, deaths, and hospitalizations due to COVID-19 was obtained from the respective government health department websites - NYC \cite{nychealth} and Chicago \cite{idphhealth}.
We considered the Chicago metro region as the ZCTAs included in 5 main counties - Cook, DuPage, Lake, Kane, and Will, totaling 253 ZCTAs.
The data for these counties were derived from the Illinois dataset which has data available starting from 18 April 2020.
The NYC health dataset spans 178 ZCTAs across the five boroughs in the NYC area, and has daily updated data starting from 3 April 2020.
The resultant dataset has missing information about tests between 18 May and 6 June, so testing rate data was not considered in this study.

\subsubsection*{Socioeconomic Factors}
The 2017 ACS was used to obtain socioeconomic variables of interest at the CBG level which were then aggregated at the ZCTA level.
A principal component analysis of these variables resulted in the selection of 6 main measures of socioeconomic standing which are listed in Table \ref{table: variables used in SEM}.

\subsection*{Mobility Metrics}
Exposure-based mobility was measured with three metrics - two social distancing metrics pertaining to residents' movement outside of their neighborhoods, and a POI-based contact exposure index (CEI).

\subsubsection*{Duration and Proportion of Stay at Home}
We used two measures of the degree of compliance to the stay-at-home orders issued in NYC and Chicago in March 2020 using the dwell time composition of mobile devices - \textit{PropHome} and \textit{TimeHome}, which are described in Table \ref{table: variables used in SEM}, considering them reasonable indirect measures of social distancing practices \cite{Jay2020}.
These metrics are measured by SafeGraph based on their estimated assignment of device owners to their home neighborhoods and provided at the CBG level, which we aggregated to the ZCTA level.
For more details about these metrics, see \href{https://docs.safegraph.com/docs/social-distancing-metrics}{https://docs.safegraph.com/docs/social-distancing-metrics}.

\subsubsection*{Contact Exposure Index (CEI)}
We define a Contact Exposure Index (CEI) that estimates the amount of exposure subjected to an individual during their visit to a place of interest (e.g. school, hospital) and possibly come in close proximity with other visitors.
It takes into account three key attributes of such trip-making - the number of people coming in contact with each other, the spacing between them at the time of contact, and the duration of this contact.

The first component is directly measured by the number of trips to a given place in a given day and has been used extensively in estimating the effect of mobility on community transmission \cite{Jia2020, Badr2020, Kraemer2020}.
The other two components require microscopic details for exact computation which are generally not available on a large scale. Hence, we approximate these by making some general assumptions on the trips to POIs.
These assumptions are that (i) visitors are uniformly spread out on the floor area of the POI at any given time, so that they can be assumed to be arranged in a square grid, (ii) all visitors arrive at the POI at the beginning of the hour of their trip, (iii) as in the worst case, every visitor comes into contact with every other visitor for the entire duration of their stay at the POI, which may vary individually.


We define contact exposure index of a given POI in a given hour as the worst-case total contact duration of its visitors divided by the square root of the POI floor square footage.
It is measured in minutes/foot.
It can be easily aggregated over higher scopes (such as POI-daily level or ZCTA-daily level) by simply summing over hours.
Its expressions for the 3 scopes considered here are given below.
\begin{equation}
    \text{POI-hourly: } E_{p, h} = \frac{\tau_{p, h}}{\sqrt{A_p}}\;;
    \qquad \text{POI-daily: } E_{p, t} = \sum_{h=1}^{24} E_{p, h}\;;
    \qquad \text{ZCTA-daily: } E_{i, t} = \sum_{p\in z} E_{p, t}
\end{equation}

Here, $\tau_{p,h}$ is the total contact duration (in minutes) of POI $p$ (lying in ZCTA $z$) during hour $h$ of day $t$ and $A_p$ is the floor area of the POI $p$ in squared feet, which excludes parking lots but may include unusable space such as for fixtures.
$\tau_{p,h}$ is the sum of contact duration of each pair of visitors in hour $h$, given by the minimum of their dwell times, since both visitors have to be physically present at the POI to come into contact with each other.

This can be illustrated with an example. Suppose we have 6 persons (say, A-F) visiting a POI between 1:00 and 2:00 PM with the following dwell times (minutes): {A:10, B:10, C:20, D:20, E:20, F:40}.
According to our assumptions, visitors A and B come into contact with everyone else for 10 minutes, so the total contact duration of both A and B is $10*5=50$ min.
Visitors C, D, E come into contact for 10 minutes with A and B and 20 minutes with everyone else, so their contact duration is $10*2+3*20=80$ min.
Finally, F contacts A and B for 10 minutes and with C, D, and E for 20 minutes, so its contact duration is $10*2+3*20=80$.
Since each pair has to be counted once, the total contact duration is half of the sum of these individuals' contact duration, i.e., $0.5*(50+50+80+80+80+80)=210$ min.

When the dwell time distribution is discrete, such as in the SafeGraph data, it has to be assumed that each trip in a given bucket has a dwell time equal to the representative point of that bucket.
For a $k$-bucketed distribution, the expression of total contact duration is
\begin{equation}
    {\tau}_{p,h} = \frac{1}{2}\sum_{i=1}^k \mu_i n_i \left(n_i - 1 + \sum_{j=i+1}^k n_j \right)
    \label{eqn: general total contact duration}
\end{equation}
Here, $n_i$ is the number of trips to POI $p$ in hour $h$ whose dwell time lies in the $i^{th}$ bucket, with $\mu_i$ denoting the representative point of that bucket.
It can be seen that visits in higher duration buckets dominate this measure, making it more realistic in terms of the compounding effect of trip duration on contact exposure.
It also follows from this expression that a higher value of $k$ (corresponding to finer intervals) provides a more accurate estimate of total contact duration.
The dwell time distribution in the SafeGraph data is given by the $k=4$ buckets: $[0, 5), [5, 20), [20, 60), [60, \infty)$, so we chose $\mu=[2.5,12.5,40,60]$ minutes.

\subsection*{Structured Equations Model}
\label{sec: sem}
The model form used in the daily SEMs is represented by the following system of equations. The variables are described in Table \ref{table: variables used in SEM}.
\begin{subequations}
\begin{align}
    y_{i,t} &= \alpha_y + \beta_0 \cdot y_{i,t-1} + \beta_{\eta} \cdot \eta_{i,w} + \epsilon_{y_{i,t}} + \mu_{y_i} + \nu_{y_t} \\
    \eta_{i,w} &= \alpha_S + \beta_S^T \mathbf{S}_i + \epsilon_{S_{i,t}} + \mu_{S_i} + \nu_{S_t} \\
    E_{i,w} &= \sum_{z=t-7}^{t-1} E_{i,z} = \alpha_E + 1\cdot\eta_{i,w} + \epsilon_{E_{i,t}} + \mu_{E_i} + \nu_{E_t} \\
    P_{i,w} &= \sum_{z=t-7}^{t-1} N_{i,z} P_{i,z} = \alpha_P + \beta_P\cdot\eta_{i,w} + \epsilon_{P_{i,t}} + \mu_{P_i} + \nu_{P_t} \\
    T_{i,w} &= \sum_{z=t-7}^{t-1} N_{i,z} T_{i,z} = \alpha_T + \beta_T\cdot\eta_{i,w} + \epsilon_{T_{i,t}} + \mu_{T_i} + \nu_{T_t}
\end{align}
\end{subequations}

For each day $t$, this form assumes a causal impact of static socioeconomic variables ($\mathbf{S}_i$, all mutually correlated) on the total latent exposure, $\eta_{i,w}$, measured by the daily mobility variables ($\mathbf{M}_{i,w}$, all mutually correlated) which itself influences the number of cases on that day, $y_{i,t}$.
The effects of unmeasured contributory factors, such as testing rate and human behavior (better hygiene, use of protective face masks, personal motivation to travel, etc.) are captured by the number of cases in the neighborhood on the previous day, $y_{i,t-1}$.
Also, we consider the total mobility of the past 7 days ($w=[t-7,\:t)$) as contributing to the growth of cases on day $t$ instead of the exposure on that day, based on a manifestation period of 7 days for COVID-19 (5 days for incubation \cite{lauer2020incubation} + 2 days for reporting).
In these equations, $\epsilon$, $\mu$, and $\nu$ respectively denote the random spatiotemporal, fixed spatial, and fixed temporal error terms.


\section*{Data Availability}
The COVID-19 cases data are available on the state department of health websites of NYC\cite{nychealth} and Chicago\cite{idphhealth}. 
The ACS (census) data are available from \href{https://www.census.gov}{www.census.gov}.
The mobility data from SafeGraph are available for free on request from \href{https://www.safegraph.com/covid-19-data-consortium}{www.safegraph.com/covid-19-data-consortium}.
The datasets generated during this study and the code used to generate them are available in the repository \href{https://www.github.com/umnilab/covid19_safegraph/}{www.github.com/umnilab/covid19\_safegraph}.

\bibliography{main}


\section*{Author Contributions}
SU and RV conceived the study design.
RV and TY analyzed the results.
RV wrote the manuscript text.
All authors reviewed the manuscript.

\section*{Additional Information}
\subsection*{Competing Interests}
The authors declare no competing interests.

\newpage


\end{document}